\documentclass[12pt,a4paper]{article}

\RequirePackage[l2tabu, orthodox]{nag}

\usepackage{mjheppub}

\usepackage{pgf}
\usepackage{tikz}
\usetikzlibrary{calc}

\title{\boldmath Liouville theory \\ with a central charge less than one}

\author{Sylvain Ribault$^1$}
\author{and Raoul Santachiara$^2$}

\affiliation{\vspace{2mm} $^1$ CEA Saclay, Institut de Physique Th\'eorique 
}
\affiliation{$^2$ Universit\'e Paris-Sud, Laboratoire de Physique Th\'eorique et Mod\`eles Statistiques}

\emailAdd{sylvain.ribault@cea.fr, raoul.santachiara@u-psud.fr}

\preprint{}

\abstract{
We determine the spectrum and correlation functions of Liouville theory with a central charge less than (or equal) one. 
This completes the definition of Liouville theory for all complex values of the central charge. 
The spectrum is always spacelike, and there is no consistent timelike Liouville theory.
We also study the non-analytic conformal field theories that exist at rational values of the central charge.  
Our claims are supported by numerical checks of crossing symmetry. 
We provide Python code for computing Virasoro conformal blocks, and correlation functions in Liouville theory and (generalized) minimal models. 
}

\keywords{Conformal field theory, Liouville theory, crossing symmetry, central charge, conformal blocks}

\begin{document}

\maketitle
\flushbottom

\section{Introduction}

The existence and properties of two-dimensional conformal field theories crucially depend on the value of the central charge $c$, which is a parameter of the Virasoro symmetry algebra. 
For example minimal models, the only consistent theories whose spectrums are made of finitely many irreducible representations, exist for the values 
\begin{align}
 c = 1 - 6\frac{(p-q)^2}{pq} \ ,
\label{cpq}
\end{align}
where $p>q\geq 2$ are integers. 
Liouville theory, the simplest nontrivial theory with a continuous spectrum, is universally believed to exist for $c\geq 25$. 
Moreover, it is widely believed that Liouville theory has a consistent analytic continuation to all values of $c$ except $c\leq 1$.
(See \cite{rib14} for a review.)
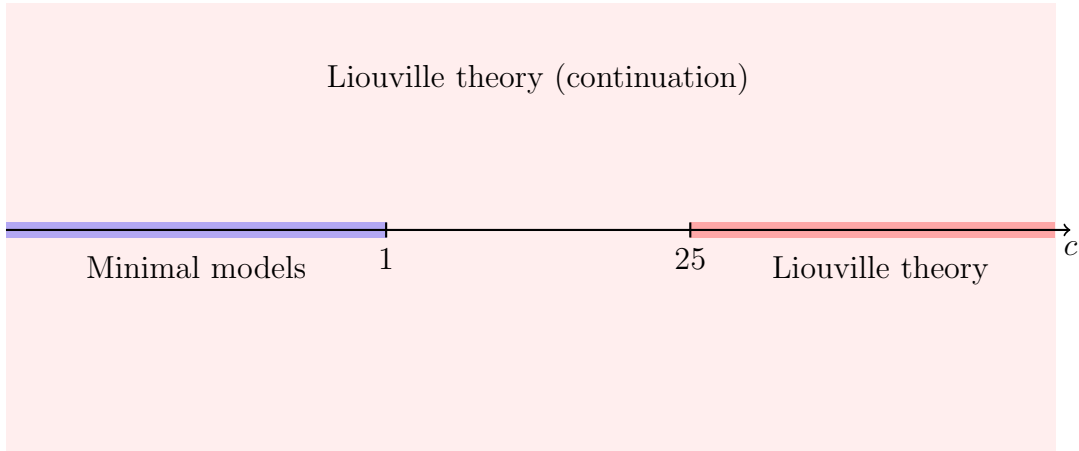
\begin{center}
 \begin{tikzpicture}
\fill[red, opacity = .07] (-7,-3) rectangle (6.8, 3);
\fill[blue, opacity = .3] (-7,-.1) rectangle (-2, .1);
\fill [red, opacity = .3] (2,-.1) rectangle (6.8,.1);
  \draw[thick, ->] (-7, 0) -- (7,0) node[below]{$c$};
  \draw[thick] (-2, -.1) node[below]{$1$} -- (-2, .1);
\draw[thick] (2, -.1) node[below]{$25$} -- (2, .1);
\node at (-4.5, -.2)[below] {Minimal models};
\node at (4.5, -.2)[below] {Liouville theory};
\node at (0, 2) {Liouville theory (continuation)};
 \end{tikzpicture}
\end{center}
For $c\leq 1$, the analytic continuation of Liouville theory does not have a well-defined limit. 
On the other hand, the crossing symmetry equations for the three-point function behave similarly for $c\leq 1$ and for $c\geq 25$: the degenerate equations, that is those of the crossing symmetry equations which can be written explicitly, have a unique solution in these cases.  
For $c\geq 25$, that unique solution is the Liouville three-point function \cite{tes95}.
For $c\leq 1$, the unique solution is known \cite{sch03, zam05, kp05a}, but it is not known whether it solves the rest of the crossing symmetry equations, and therefore corresponds to a consistent theory. 

Recent findings on the two-dimensional Potts model provide renewed motivation for studying Liouville theory with $c\leq 1$.
In its formulation in terms of random clusters,
the $Q$-states Potts model comes with a parameter $1\leq Q\leq 4$.
Special cases include percolation ($Q=1$) and the Ising model ($Q=2$).
In the critical limit, the clusters become conformally invariant random fractals, and the 
parameter $Q$ is related to the central charge $c$ with $0\leq c\leq 1$. (See \cite{du06} for a review.)
While the critical exponents of the $Q$-states Potts model have been known for a long time, not much is known on the correlation functions. 
Recently, it was observed that probabilities for three points to belong to the same cluster are given by certain values of the  three-point function which solves the degenerate crossing symmetry equations \cite{devi11, ziff11, psvd13}. 
This suggests that this three-point function should belong to a consistent theory. 
Our aim is to propose a precise definition of that theory.

The crossing symmetry equations involve not only the three-point structure function, but also the spectrum of the theory. 
So we should determine the spectrum of Liouville theory with $c\leq 1$. 
It was suggested that there exists a timelike Liouville theory \cite{st03}, where conformal dimensions are bounded from above. 
We will dismiss this as inconsistent, and focus on the hypothesis that conformal dimensions are bounded from below, as is the case for the other values of $c$. 
With this hypothesis, correlation functions appear to be singular, but the singularities can be eliminated by a simple regularization. 

The crucial evidence that our ideas are correct will be provided by numerical checks of crossing symmetry. 
For that evidence to be acceptable, the corresponding computer code should be publicly available.
One could even argue that the software which is used should be free and open source. 
So we wrote Python code in the IPython Notebook format, and released it on GitHub \cite{b2P}.
In particular, the code for generating our data tables and plots can be found in the notebook ``\texttt{article\_support.ipynb}'', which has the same structure as the present article. 

Using our code, we checked crossing symmetry not only in Liouville theory with $c\leq 1$, but also in Liouville theory with generic complex values of $c$, and in generalized minimal models. 
We also checked crossing symmetry in certain non-rational theories which exist at rational values of $c$ \eqref{cpq}, and which generalize the $c=1$ Runkel-Watts theory.
And we tested a few speculations of our own on the existence of further theories at rational values of $c$.

\section{Liouville theory in the conformal bootstrap approach}

Let us sketch the conformal bootstrap approach to Liouville theory, and explain how the results depend on the value of $c$. More details and references can be found in the review article \cite{rib14}.

\subsection{Definition and spectrum of Liouville theory}

We define Liouville theory as a two-dimensional conformal field theory which obeys two main assumptions:
\begin{enumerate}
 \item The spectrum contains continuously many Verma modules of the Virasoro algebra, each of them having multiplicity one.
\item The theory depends smoothly on the central charge $c$, moreover correlation functions depend smoothly on the conformal dimensions of the fields.
\end{enumerate}
Let us spell out the meaning and first consequences of these assumptions. 
The spectrum is a representation of the symmetry algebra $\mathfrak{V}\times \bar{\mathfrak{V}}$, which is made of two copies of the Virasoro algebra $\mathfrak{V}$ with the same central charge $c$ -- these copies are called left-moving or holomorphic for $\mathfrak{V}$, and right-moving or anti-holomorphic for $\bar{\mathfrak{V}}$.
The spectrum is built from Verma modules $\mathcal{V}_\Delta$ of the Virasoro algebra, which depend on the conformal dimension $\Delta$. 
Our assumptions imply that the spectrum is diagonal,
\begin{align}
\mathcal{S} = \int d\Delta\, \mathcal{V}_\Delta \otimes \bar{\mathcal{V}}_{\Delta} \ .
\end{align}
We still have to specify the allowed values of the conformal dimension $\Delta$. 
For $c> 1$, these values can be guessed from some features of the representation theory of the Virasoro algebra, which we now review. 

The set of representations of the Virasoro algebra at a given value of $c$ is a ring for an associative product called the fusion product, which is useful in conformal field theory because it constrains the operator product expansions of the corresponding fields. 
Verma modules have the rather trivial fusion product
\begin{align}
 \mathcal{V}_{\Delta_1}\times \mathcal{V}_{\Delta_2} = \int d\Delta\, \mathcal{V}_\Delta\ .
\end{align}
However, there exist degenerate representations $\{\mathcal{R}_{(r,s)}\}_{r,s\in\mathbb{N}^*}$, whose fusion products involve only finitely many terms. 
These fusion products have simple expressions provided we parametrize the central charge $c$ in terms of the background charge $Q$ and coupling constant $b$,
\begin{align}
 c = 1 + 6Q^2 \quad , \quad Q = b+b^{-1}\ ,
\end{align}
and the conformal dimension $\Delta$ in terms of the momentum $\alpha$,
\begin{align}
 \Delta = \alpha(Q-\alpha)\ .
\end{align}
The fusion product of the degenerate representation $\mathcal{R}_{(r,s)}$ with a Verma module is a sum of $rs$ Verma modules, 
\begin{align}
 \mathcal{R}_{(r,s)} \times \mathcal{V}_\alpha = \sum_{i=\frac{1-r}{2}}^{\frac{r-1}{2}} \sum_{j=\frac{1-s}{2}}^{\frac{s-1}{2}} \mathcal{V}_{\alpha + ib + j b^{-1}}\ ,
\label{rvsv}
\end{align}
where the sums run in steps of $1$.
Algebraically, a degenerate representation is the quotient of a reducible Verma module by a nontrivial submodule,
\begin{align}
 \mathcal{R}_{(r,s)} = \frac{\mathcal{V}_{\alpha_{(r,s)}}}{\mathcal{V}_{\alpha_{(-r,s)}}}\ , 
\label{rvv}
\end{align}
where we introduce 
\begin{align}
 \alpha_{(r,s)} = \frac{1-r}{2} b + \frac{1-s}{2} b^{-1}\ .
\label{ars}
\end{align}
To conclude on Virasoro representations, let us mention that unitary Verma modules exist only if $c\geq 1$. 
More precisely, if $c>1$ then $\mathcal{V}_\Delta$ is unitary if and only if $\Delta>0$.

For $c>1$, we will use the assumption of unitarity for guiding us in guessing the spectrum. 
A first guess would be to include all Verma modules which obey the unitarity bound $\Delta >0$. 
However, we have also learned that the natural variable of Virasoro representation theory is not the conformal dimension $\Delta$, but the momentum $\alpha$, and our assumption of smoothness of correlation functions really means analyticity in $\alpha$.  
In terms of $\alpha$, the unitarity bound becomes 
\begin{align}
 \Delta > 0 \quad \Leftrightarrow \quad \alpha \in (0, Q) \cup \left(\frac{Q}{2} + i \mathbb{R}\right)\ .
\end{align}
It would be difficult for correlation functions to be analytic if $\alpha$ was allowed to live 
in the segment $(0, Q)$ with its finite endpoints. 
So we exclude this segment from the spectrum, and the allowed values of the momentum become 
\begin{align}
 \Delta \in \frac{c-1}{24} + \mathbb{R}_+ \quad \Leftrightarrow \quad \alpha \in \frac{Q}{2} + i \mathbb{R}\ .
\label{alin}
\end{align}
We will later see that this is true not only for $c>1$, but for all $c\in \mathbb{C}$. For the moment let us introduce a real variable $P$ for the momentum, which is related to $\Delta$ and $\alpha$ by
\begin{align}
 \Delta = \frac{Q^2}{4} + P^2 \quad , \quad \alpha = \frac{Q}{2} + iP\ .
\label{Pdef}
\end{align}

\subsection{Crossing symmetry equations for correlation functions}\label{sec:cse}

Solving Liouville theory means computing its correlation functions. 
An $N$-point correlation function is associated to $N$ fields, where a field is an operator-valued function on the Euclidean plane.
There is a state-field correspondence between states in the spectrum, and fields.
In a Verma module $\mathcal{V}_\alpha$, all states are obtained by acting with creation operators on the module's primary state. 
We call $V_\alpha$ (or equivalently $V_{\Delta}$) the primary field which corresponds to that primary state. 
Then the $N$-point function of primary fields
\begin{align}
 \left< \prod_{i=1}^N V_{\alpha_i}(z_i) \right> \ ,
\end{align}
is a function of the central charge, the momentums $\alpha_i$, and the complex coordinates $z_i$ on the Euclidean plane.
The fields which are obtained by acting with creation operators are called descendent fields, and their correlation functions are deduced from correlation functions of primary fields through Ward identities.

In principle, correlation functions only involve fields which correspond to states in the spectrum, whose momentums take values as in eq. \eqref{alin}. 
However, it is fruitful to consider fields which correspond to more general states, and in particular to states in degenerate representations. 
So we assume that there exists a degenerate primary field $V_{(r,s)}$ which corresponds to the primary state of the degenerate representation $\mathcal{R}_{(r,s)}$. (We are not assuming that $V_{(r,s)}$ can be obtained from $V_\alpha$ by analytic continuation, see the discussion in Section \ref{sec:tp}.)
The existence of a vanishing descendent in the representation $\mathcal{R}_{(r,s)}$ \eqref{rvv} implies the existence of a vanishing descendent field for $V_{(r,s)}$. This in turn implies that a correlation function involving $V_{(r,s)}$ obeys a Belavin--Polyakov--Zamolodchikov differential equation of order $rs$, whose solutions correspond to the $rs$ terms in the fusion product $\mathcal{R}_{(r,s)} \times \mathcal{V}_\alpha$ \eqref{rvsv}.

The main axiom which underlies the calculation of correlation functions is the existence of an associative operator product expansion,
\begin{align}
 V_{\alpha_1}(z_1) V_{\alpha_2}(z_2) 
= \int_{\frac{Q}{2}+i\mathbb{R}} d\alpha\ C(\alpha_1,\alpha_2,Q-\alpha)
\sum_{\mathcal{L}, \bar{\mathcal{L}}} \Big|g^{\mathcal{L}}_{\alpha_1,\alpha_2|\alpha}(z_1,z_2)\Big|^2\, \mathcal{L}\bar{\mathcal{L}}V_{\alpha}(z_2)\ . 
\end{align}
This expresses the product of two fields $V_{\alpha_1}(z_1) V_{\alpha_2}(z_2) $ as a linear combination of the primary field $V_\alpha$ (obtained for $\mathcal{L} = \bar{\mathcal{L}}= \mathbf{1}$) and its 
descendent fields $\mathcal{L}\bar{\mathcal{L}}V_{\alpha}$, where $\mathcal{L}$ is a creation operator in the universal enveloping algebra of the left-moving Virasoro algebra.
The coefficients in this linear combination are factorized into the universal factors  $g^{\mathcal{L}}_{\alpha_1,\alpha_2|\alpha}(z_1,z_2)$, which are determined by conformal symmetry and normalized so that $g^{\mathcal{L}=\mathbf{1}}_{\alpha_1,\alpha_2|\alpha}(z_1,z_2)=1$, and the three-point structure constant $C(\alpha_1,\alpha_2,Q-\alpha)$.
Here and in the following, the notation $\left|g^{\mathcal{L}}_{\alpha_1,\alpha_2|\alpha}(z_1,z_2)\right|^2$ does not stand for the complex modulus squared, but for a product of left-moving and right-moving quantities, namely $\left|g^{\mathcal{L}}_{\alpha_1,\alpha_2|\alpha}(z_1,z_2)\right|^2 =g^{\mathcal{L}}_{\alpha_1,\alpha_2|\alpha}(z_1,z_2) g^{\bar{\mathcal{L}}}_{\alpha_1,\alpha_2|\alpha}(\bar{z}_1,\bar{z}_2) $.

Repeatedly using the operator product expansion, any correlation function can be written as a combination of three-point structure constants, and universal quantities. 
For example, a four-point correlation function can be written as
\begin{align}
\left< \prod_{i=1}^4 V_{\alpha_i}(z_i) \right> 
= \int_{\frac{Q}{2}+i\mathbb{R}} d\alpha_s\ C(\alpha_1,\alpha_2,Q-\alpha_s) C(\alpha_s,\alpha_3,\alpha_4)\, 
\Big|\mathcal{F}^{(s)}_{\alpha_s}(\alpha_i|z_i)\Big|^2\ ,
\label{sfour}
\end{align}
where we introduced the four-point, $s$-channel conformal blocks $\mathcal{F}^{(s)}_{\alpha_s}(\alpha_i|z_i)$ -- combinations of $g^{\mathcal{L}}_{\alpha_1,\alpha_2|\alpha}(z_1,z_2)$ and similar universal quantities, summed over $\mathcal{L}$. 
Such conformal blocks are determined by conformal symmetry and therefore in principle known, just like characters of representations. 
The calculation of conformal blocks is discussed in Section \ref{sec:ccb}. 
Here we only need the following properties:
\begin{itemize}
 \item $\mathcal{F}^{(s)}_\alpha(\alpha_i|z_i)$ is a meromorphic function of $b$ and $\alpha$, whose $b$-poles occur for the ``minimal'' values $b^2\in \mathbb{Q}_{<0}$ \eqref{cpq}, and whose $\alpha$-poles occur for ``degenerate'' values $\alpha = \alpha_{(r,s)}$ \eqref{ars}.

\item For large values of $\Delta = \alpha(Q-\alpha)$, conformal blocks behave as 
\begin{align}
 \mathcal{F}^{(s)}_\alpha(\alpha_i|z_i) \underset{|\Delta| \to \infty}{\sim} (16q)^{\Delta}\quad \text{with} \quad |q| < 1\ ,
\label{dti}
\end{align}
where $q$ is a function of $z_1,\cdots z_4$.
\end{itemize}

Our decomposition of a four-point correlation function was obtained using the operator product expansion of $V_{\alpha_1}V_{\alpha_2}$, but we could alternatively use the expansion of $V_{\alpha_2}V_{\alpha_3}$, and obtain the $t$-channel decomposition,
\begin{align}
\left< \prod_{i=1}^4 V_{\alpha_i}(z_i) \right> 
= \int_{\frac{Q}{2}+i\mathbb{R}} d\alpha_t\ C(\alpha_2,\alpha_3,Q-\alpha_t) C(\alpha_t,\alpha_4,\alpha_1)\, 
\Big|\mathcal{F}^{(t)}_{\alpha_t}(\alpha_i|z_i)\Big|^2\ .
\end{align}
The equality of the $s$- and $t$-channel decompositions is called crossing symmetry, and is equivalent to the associativity of the operator product expansion. (In particular, no further constraints arise from the $u$-channel decomposition \cite{zz95}.) 
Crossing symmetry can be represented using the following graphs, where each vertex corresponds to a three-point structure constant:
\begin{equation}
 \begin{tikzpicture}[baseline=(current  bounding  box.center), very thick, scale = .35]
\draw (-1,2) node [left] {$\alpha_1$} -- (0,0) -- node [above] {$\alpha_s$} (4,0) -- (5,2) node [right] {$\alpha_4$};
\draw (-1,-2) node [left] {$\alpha_2$} -- (0,0);
\draw (4,0) -- (5,-2) node [right] {$\alpha_3$};
\draw (16,3) node [left] {$\alpha_1$} -- (18,2) -- node [left] {$\alpha_t$} (18,-2) -- (16, -3) node [left] {$\alpha_2$};
\draw (18,2) -- (20,3) node [right] {$\alpha_4$};
\draw (18,-2) -- (20, -3) node [right] {$\alpha_3$};
\node at (-5,0) {$\displaystyle \int d\alpha_s$};
\node at (11,0) {$ = \quad \displaystyle \int d\alpha_t$};
 \end{tikzpicture}
\end{equation}
Crossing symmetry provides an unwieldy set of equations for the three-point structure constant $C(\alpha_1,\alpha_2,\alpha_3)$: we have as many equations as choices of values of $\alpha_1,\cdots \alpha_4$, each equation is quadratic, and has infinitely many terms. 
Simpler equations, with finitely many terms, can be obtained if we replace one field, say $V_{\alpha_4}$, with a degenerate field $V_{(r,s)}$. 
Then $\alpha_s$ and $\alpha_t$ take $rs$ possible values, which are determined by the fusion rules \eqref{rvsv}. 
These values are in general outside $\frac{Q}{2}+i\mathbb{R}$, which is not a problem since correlation functions are analytic in $\alpha$.
The simplest nontrivial examples are obtained for $(r,s) \in \{(1,2),(2,1)\}$. 
With $V_{(2,1)}$, we obtain the following equation:
\begin{equation}
\sum_\pm^{\phantom{\pm}} \ \
 \begin{tikzpicture}[baseline=(current  bounding  box.center), very thick, scale = .35]
\draw (-1,2) node [left] {$\alpha_1$} -- (0,0) -- node [above] {$\alpha_3\pm\frac{b}{2}$} (4,0) -- (5,2) node [right] {$(2,1)$};
\draw (-1,-2) node [left] {$\alpha_2$} -- (0,0);
\draw (4,0) -- (5,-2) node [right] {$\alpha_3$};
\end{tikzpicture}
\ \ = \sum_\pm^{\phantom{\pm}} \quad 
\begin{tikzpicture}[baseline=(current  bounding  box.center), very thick, scale = .35]
\draw (16,3) node [left] {$\alpha_1$} -- (18,2) -- node [left] {$\alpha_1\pm\frac{b}{2}$} (18,-2) -- (16, -3) node [left] {$\alpha_2$};
\draw (18,2) -- (20,3) node [right] {$(2,1)$};
\draw (18,-2) -- (20, -3) node [right] {$\alpha_3$};
 \end{tikzpicture}
 \label{stgraph}
\end{equation}
This leads to relations involving $C(\alpha_1,\alpha_2,\alpha_3\pm\frac{b}{2})$ and $C(\alpha_1\pm\frac{b}{2},\alpha_2,\alpha_3)$, which actually imply
\begin{equation}
 \frac{C(\alpha_1,\alpha_2,\alpha_3+b)}{C(\alpha_1,\alpha_2,\alpha_3)} 
=  \frac{b^{-2bQ}\gamma(2b\alpha_3)\gamma(2b\alpha_3+b^2)}{\prod_{\pm,\pm}\gamma(b\alpha_3\pm b(\alpha_1-\frac{Q}{2}) \pm b(\alpha_2-\frac{Q}{2}))}\ ,
\label{ccg}
\end{equation}
where the function $\gamma(x)$ is defined from Euler's Gamma function by
\begin{equation}
\gamma(x) = \frac{\Gamma(x)}{\Gamma(1-x)}\ .
\end{equation}
If we had used the degenerate field $V_{(1,2)}$ instead of $V_{(2,1)}$, the equation we would have obtained would be related to eq. \eqref{ccg} by $b\rightarrow b^{-1}$. 
Summing up, we have two equations of the type
\begin{equation}
 \frac{C(\alpha_1,\alpha_2,\alpha_3+b)}{C(\alpha_1,\alpha_2,\alpha_3)}  = \text{known} \ , \quad \frac{C(\alpha_1,\alpha_2,\alpha_3+b^{-1})}{C(\alpha_1,\alpha_2,\alpha_3)}  = \text{known} \ ,
\label{cck}
\end{equation}
which we call degenerate crossing symmetry equations.
(These equations appear to be linear, whereas crossing symmetry equations should be quadratic in the three-point structure constant. 
Actually, degenerate values of the structure constant of the type of $C((1,2), \alpha, \alpha\pm\frac{b}{2})$ are hidden in the right-hand sides.)

\subsection{Solving the crossing symmetry equations}\label{sec:scse}

The degenerate crossing symmetry equations determine how the three-point structure constant behaves under shifts of momentums by $b$ and $b^{-1}$.  
Let us represent $b$ and $b^{-1}$ as vectors in the complex plane:
\begin{equation}
 \begin{tikzpicture}[baseline=(current  bounding  box.center)]
\draw (0, 0) node[left]{$i$} -- (0, -1) node[below left] {$0$} -- (1, -1) node[below] {$1$};
\draw [red, ultra thick, <->] (4,3) -- (4,1) node[fill, circle, minimum size = 2mm, inner sep = 0]{} -- (4,-.3);
\draw [red, ultra thick, <->] (8,3) -- (7,1) node[fill, circle, minimum size = 2mm, inner sep = 0]{}-- (7.6,-.2);
\draw [red, ultra thick, <->] (12,1) -- (10,1) node[fill, circle, minimum size = 2mm, inner sep = 0]{} -- (11.3,1) ;
\node at (4, -1){$\begin{array}{c} b\in i\mathbb{R} \\ c\leq 1 \end{array}$};
\node at (7.5, -1){$\begin{array}{c} b\in \mathbb{C} \\ c\in\mathbb{C} \end{array}$};
\node at (11, -1){$\begin{array}{c} b\in \mathbb{R} \\ c\geq 25 \end{array}$};
 \end{tikzpicture}
\end{equation}
Here and in the following, $b\in\mathbb{C}$ should be understood as $b\notin \mathbb{R}\cup i\mathbb{R}$, and similarly $c\in \mathbb{C}$ should be understood as $c\notin (-\infty, 1)\cup (25,\infty)$.
For a generic value of $b\in \mathbb{R}\cup i\mathbb{R}$, the vectors $b$ and $b^{-1}$ are aligned and incommensurable.
Therefore, there exists a unique smooth solution of the shift equations, up to an overall momentum-independent factor. 
We will call this solution $C^{\text{DOZZ}}$ after Dorn--Otto and Zamolodchikov--Zamolodchikov if $c\geq 25$, and $C^{c\leq 1}$ if $c \leq 1$. 
For $b\in\mathbb{C}$, the vectors $b$ and $b^{-1}$ are not aligned, and there exist non-constant smooth functions which are periodic in both $b$ and $b^{-1}$. 
Multiplying a solution with such smooth functions yields more solutions. 

A three-point structure constant must solve not only the degenerate crossing symmetry equations, but also the non-degenerate equations.
If $c\leq 1$ or $c\geq 25$, uniqueness of the solution of the degenerate equations suggests that it has good chances of solving all equations. 
Moreover, for $c\geq 25$, Liouville theory is believed to exist and to be consistent, based in particular on semi-classical approaches near the $c\to +\infty$ limit.
On the other hand, for $c\in\mathbb{C}$, it would be very implausible that each one of the infinitely many solutions of the degenerate equations solves all crossing symmetry equations. 
The existence of more than one solution of these vastly overdetermined equations would already be implausible. 

We will now describe the three-point constant $C^\text{DOZZ}$ for $c\geq 25$, and explain how it also provides a solution of Liouville theory for $c\in\mathbb{C}$.
To solve the degenerate crossing symmetry equations, we manifestly need functions whose shifts by both $b$ and $b^{-1}$ generate Gamma functions as in eq. \eqref{ccg}. 
And indeed there exists a function $\Upsilon_b(x)$ such that 
\begin{align}
  \frac{\Upsilon_b(x+b)}{\Upsilon_b(x)} = b^{1-2bx} \gamma(bx)\quad \text{and} \quad \frac{\Upsilon_b(x+\frac{1}{b})}{\Upsilon_b(x)} = b^{\frac{2x}{b}-1} \gamma(\tfrac{x}{b})\ .
\label{upup}
\end{align}
This function can be defined either by an integral formula, which is valid for $0<\Re x<\Re Q$, 
\begin{align}
 \log\Upsilon_b(x) = \int_0^\infty \frac{dt}{t} \left[\left(\tfrac{Q}{2}-x\right)^2 e^{-2t} -\frac{\sinh^2\left(\frac{Q}{2}-x\right)t}{\sinh bt\sinh\frac{t}{b}}\right]\ ,
\label{lup}
\end{align}
or by a product formula, which is valid for all $x\in\mathbb{C}$, 
\begin{align}
 \Upsilon_b(x) = \lambda_b^{(\frac{Q}{2}-x)^2}\prod_{m,n=0}^\infty f\left(\frac{\frac{Q}{2}-x}{\frac{Q}{2}+mb+nb^{-1}}\right) \quad \text{with} \quad f(x)=(1-x^2)e^{x^2}\ ,
\label{upp}
\end{align}
where $\lambda_b$ is an unimportant constant.
The product formula makes it clear that $\Upsilon_b(x)$ is analytic on $\mathbb{C}$, with infinitely many simple zeros:
\begin{align}
 \Upsilon_b(x) = 0 \quad \Leftrightarrow \quad x \in \left(-b\mathbb{N}-b^{-1}\mathbb{N}\right) \cup \left(Q + b\mathbb{N} + b^{-1} \mathbb{N}\right)\ .
\end{align}
And the integral formula allows us to compute the large momentum behaviour
\begin{align}
 \log \Upsilon_b(\tfrac{Q}{2}  + iP) \underset{P\to \infty}{=} -P^2\log|P| +\frac32 P^2 + o(P^2)\ .
\label{lupi}
\end{align}
The function $\Upsilon_b$ is well-defined except for $b\in i\mathbb{R}$, where the product and integral formulas diverge. 

Using the function $\Upsilon_b(x)$, we now write the three-point structure constant 
in terms of the variable $P$ \eqref{Pdef}, 
\begin{align}
 C^{\text{DOZZ}}(P_1,P_2,P_3) 
= \frac{\prod_{i=1}^3\Upsilon_b(Q+2iP_i) }{\prod_{\pm,\pm}\Upsilon_b\left(\frac{Q}{2}+i(P_1\pm P_2\pm P_3)\right)}\ . 
\label{cdozz}
\end{align}
The symmetry of $C^{\text{DOZZ}}(P_1,P_2,P_3)$ under permutations of $\{P_i\}$ follows from the identity $\Upsilon_b(x) = \Upsilon_b(Q-x)$. 

Like the function $\Upsilon_b$, this three-point structure constant is well-defined for all $c$ except $c\leq 1$. 
A four-point function can then be computed using its $s$-channel decomposition \eqref{sfour}. 
Let us check that the integral over $P_s\in\mathbb{R}$ converges. 
To begin with, the poles of the integrand are safely away from the integration line. 
There are indeed poles from the conformal blocks at 
\begin{align}
 P_s \in \bigcup_\pm\, \pm \frac{i}{2}\left(Q + b\mathbb{N} +b^{-1}\mathbb{N}\right)\ , 
\label{polesblocks}
\end{align}
and poles from the structure constants at 
\begin{multline}
 P_s \in \bigcup_{\pm,\pm,\pm} \Big(\pm P_1 \pm P_2 \pm i\left(\tfrac{Q}{2} + b\mathbb{N} +b^{-1}\mathbb{N}\right) \Big)
 \\
\bigcup \bigcup_{\pm, \pm, \pm} \Big(\pm P_3 \pm P_4 \pm i\left(\tfrac{Q}{2} + b\mathbb{N} +b^{-1}\mathbb{N}\right) \Big)\ . 
\label{polessc}
\end{multline}
Drawing a discrete set of poles $P + i(b\mathbb{N} + b^{-1}\mathbb{N})$ as a wedge,
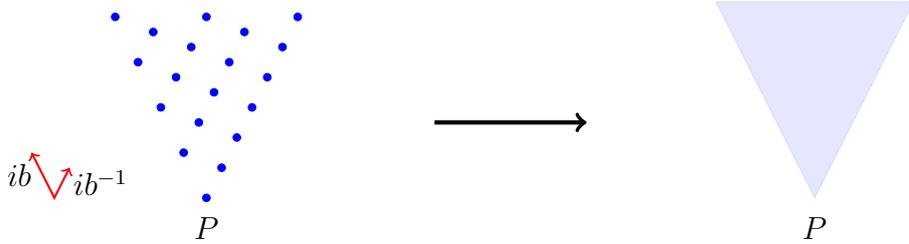
\begin{center}
 \begin{tikzpicture}
\node at (0, -.4) {$P$};
\node at (8, -.4) {$P$};
\draw [ultra thick, ->] (3, 1) -- (5, 1);
\filldraw[blue, opacity = .1] (8,0) -- (6.7, 2.6) -- (9.3, 2.6) -- cycle;
\draw [red, thick, <->] (-2.3, .6) -- node[left, black]{$ib$} (-2,0) -- node[right, black]{$ib^{-1}$} (-1.8, .4);
\clip (-1.4, -.2) -- (1.4, -.2) -- (1.4, 2.5) -- (-1.4, 2.5) -- cycle;
\foreach \x in {0, 1,...,4}{
  \foreach \y in {0, 1,...,6}{
    \node[draw,circle,inner sep=1pt,fill,blue] at (-.3*\x +.2*\y, .6*\x +.4*\y) {};
  }
}
 \end{tikzpicture}
\end{center}
the poles of the integrand form $9$ wedges on each side of the integration line, leaving a strip of width $\Re Q$ free of poles:
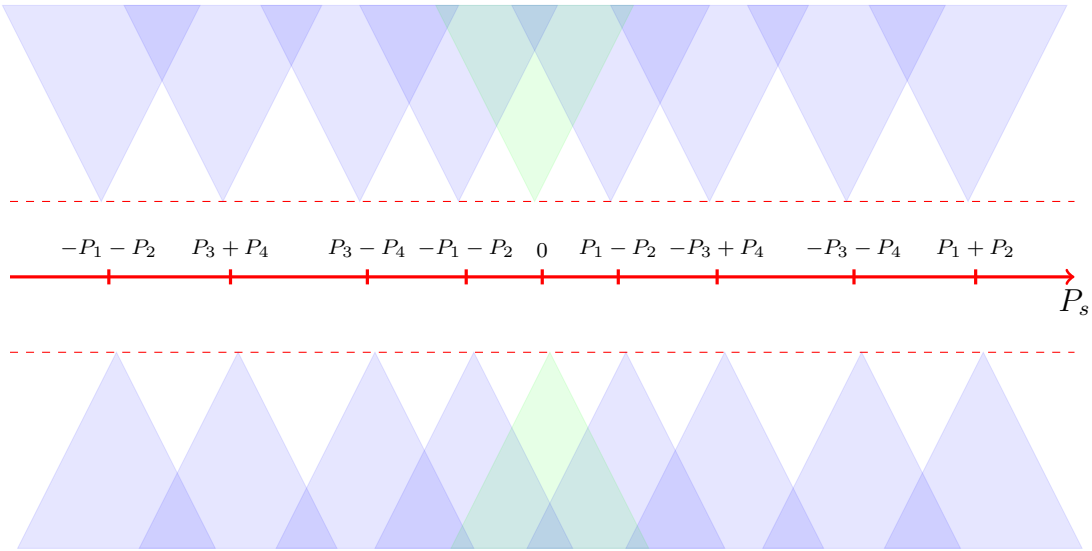
\begin{center}
 \begin{tikzpicture}
\newcommand{\polewedge}{\filldraw[opacity = .1] (0,0) -- (-1.3, 2.6) -- (1.3, 2.6) -- cycle;}
\newcommand{\twowedges}[3][]{
\begin{scope}[blue, #1, shift = {(#2 -.1, 1)}] \polewedge \end{scope}
\begin{scope}[blue, #1, shift = {(#2 + .1, -1)}, rotate = 180] \polewedge \end{scope}
\draw[red, very thick] (#2, -.1) -- (#2, .1) node[above, black]{\scriptsize{#3}};
}
\draw[red, very thick, ->] (-7, 0) -- (7, 0) node[below, black]{$P_s$};
\draw[red, dashed] (-7,1) -- (7,1);
\draw[red, dashed] (-7,-1) -- (7,-1);
\twowedges{5.7}{$P_1+P_2$}
\twowedges{1}{$P_1-P_2$}
\twowedges{-1}{$-P_1-P_2$}
\twowedges{-5.7}{$-P_1-P_2$}
\twowedges{-4.1}{$P_3+P_4$}
\twowedges{-2.3}{$P_3-P_4$}
\twowedges{2.3}{$-P_3+P_4$}
\twowedges{4.1}{$-P_3-P_4$}
\twowedges[green]{0}{$0$}
 \end{tikzpicture}
\end{center}
It remains to be checked that the four-point function, as an integral over $P_s$, converges at $P_s = \pm \infty$. 
Using the behaviour of the conformal blocks \eqref{dti}, and that of the three-point structure constants, which follows from eq. \eqref{lupi}, 
we find that the integrand behaves as 
\begin{align}
 C^\text{DOZZ}(P_1,P_2,P_s)C^\text{DOZZ}(-P_s,P_3,P_4) \left|\mathcal{F}^{(s)}_{P_s}(P_i|z_i)\right|^2 \underset{P_s\to \pm\infty}{\sim} |q|^{2P_s^2}\ ,
\label{pinf}
\end{align}
with $|q|<1$. So the integral converges. This means that the four-point function is well-defined, as computed from its $s$-channel decomposition. 
The same holds for the $t$-channel decomposition. 
Assuming that these two decompositions agree for $c\leq 25$, they must agree for all values of $c$ except $c\leq 1$, because of the $b$-independent convergent behaviour of the integrals, and the fact that the integrands are analytic in $b$.
In other words, the DOZZ three-point structure constant obeys crossing symmetry wherever it exists. (See also the discussion in \cite{hmw11}.)

We checked this numerically for various values of the parameters. 
For example, with $(P_i) = (1.3, 1.01, 0.45, 0.22)$ and $(z_i) = (0.26, 0, \infty, 1)$, 
we find the following values for the real parts of the $s$- and $t$-channel decompositions, depending on the central charge $c$:
\begin{align}
\renewcommand*{\arraystretch}{1.1}
 \begin{array}{|r|r|r|r|} \hline c & s\text{-channel} & t\text{-channel} & \text{precision} \\ \hline  \hline 36.74 & 256250.53844 & 256245.59396 & 1.9 \times 10^{-5} \\ \hline 17.55 & 1104.88358 & 1104.85666 & 2.4 \times 10^{-5} \\ \hline 2.12 & 2.04664 & 2.04659 & 2.6 \times 10^{-5} \\ \hline 3.00+4.00i & -2.97769 & -2.97757 & 3.6 \times 10^{-5} \\ \hline 0.50+2.00i & -5.92002 & -5.91968 & 4.3 \times 10^{-5} \\ \hline -1.00+2.10i & -11.43323 & -11.43175 & 8.3 \times 10^{-5} \\ \hline  \end{array} 
\end{align}
Details such as the normalizations of the four-point functions can be found in the code. 
What matters is the agreement between both channels, up to a small relative difference which we call the precision. 
That difference can be made smaller by increasing the numerical accuracy of the computations, see the Appendix.

\section{The problem with \texorpdfstring{$c\leq 1$}{c<=1}}

Having reviewed the definition and solution of Liouville theory for generic complex values of the central charge, we now turn to the case $c\leq 1$. 

\subsection{The three-point structure constant} \label{sec:tp}

In the case $c\leq 1$, the 
degenerate crossing symmetry equations \eqref{cck} have the unique smooth solution \cite{sch03, zam05, kp05a}
\begin{align}
C^{c\leq 1}(P_1, P_2, P_3) = \frac{\prod_{\pm, \pm} \Upsilon_\beta\Big(\frac12(\beta+\frac{1}{\beta}) + P_1 \pm P_2 \pm P_3\Big)}{\prod_{i=1}^3 \Upsilon_\beta(\frac{1}{\beta} +2P_i)}\ ,
\label{cleq1}
\end{align}
where we introduce the parameter 
\begin{align}
 \beta = -ib \in \mathbb{R}\ .
\end{align}
This solution has sometimes been rejected as unphysical \cite{sch03}, because in the limit $P_1\to i\frac{Q}{2}=\frac12(\beta^{-1}-\beta)$ the resulting three-point function does not vanish for $P_2\neq \pm P_3$, and therefore cannot be interpreted as a two-point function. 
This means that the field $\underset{\alpha\to 0}{\lim} V_{\alpha}$ differs from the degenerate field $V_{(1,1)}\propto \text{identity}$, which corresponds to the primary state of the irreducible coset $\mathcal{R}_{(1,1)}$ \eqref{rvv}.
Instead, $\underset{\alpha\to 0}{\lim} V_{\alpha}$
corresponds to the primary state of the reducible Verma module $\mathcal{V}_{\alpha_{(1,1)}=0}$.
So we have a non-degenerate field with zero conformal dimension, which however has nothing to do with logarithmic conformal field theory, since the Virasoro generator $L_0$ is diagonalizable in the reducible Verma module $\mathcal{V}_0$ (as in all Verma modules). 
This contrasts with Liouville theory with generic $c$, where $\underset{\alpha\to 0}{\lim}V_{\alpha}$ happens to be degenerate, due to the analytic properties of the DOZZ three-point structure constant.
Nevertheless, as already pointed out in \cite{hmw11}, there is no reason to expect that $\underset{\alpha\to 0}{\lim} V_{\alpha}$ should always be degenerate.

The structure constant $C^{c\leq 1}$ is a meromorphic function of the momentums $P_i$, with countably many poles. 
In particular, the purely imaginary values of the momentums do not play any distinguished role, and there is nothing intrinsically timelike about $C^{c\leq 1}$, although it has sometimes been called the timelike Liouville three-point function \cite{psvd13}. 
Our notation $C^{c\leq 1}$ instead emphasizes the domain where it is the unique solution of the degenerate crossing symmetry equations. 

The structure constant $C^{c\leq 1}$ obeys the following equation, which involves two different values 
of the parameter $\beta$,
\begin{align}
\frac{16^{-2P_s^2}C_{(\sqrt{2}\beta)}^{c\leq 1} \left( \sqrt{2}P, \frac{P_s}{\sqrt{2}}, -\frac{P_s}{\sqrt{2}} \right) }
{
C_{(\beta)}^{c\leq 1} \left( P,\frac{1}{4\beta},P_s \right)
C_{(\beta)}^{c\leq 1} \left( -P_s,\frac{1}{4\beta},-\frac{1}{4\beta} \right) 
}
=
\nu_{\beta}2^{P(3\beta-\frac{1}{\beta}-4P)} 
\frac{\Upsilon_{\sqrt{2}\beta}\left(\sqrt{2}(P+\frac{\beta}{2}+\frac{1}{4\beta})\right)}{\Upsilon_{\sqrt{2}\beta}\left(\sqrt{2}(P+\frac{1}{4\beta})\right)}\ ,
\label{cccg}
\end{align}
where $\nu_\beta$ is a function of $\beta$. 
This relation is similar to a relation for the DOZZ structure constant, which is an important ingredient in proving modular invariance of Liouville theory on the torus \cite{hjs09}. 
The other ingredient of the proof is a relation for conformal blocks, which is true for all values of the central charge by analyticity of the conformal blocks. 
The proof then relates modular invariance on the torus to crossing symmetry on the sphere -- the structure constants in eq. \eqref{cccg} are relevant for a torus one-point function  (numerator) and sphere four-point function (denominator):
\begin{align}
 \begin{tikzpicture}[baseline=(current  bounding  box.center), very thick, scale = .35]
\draw (-1,2) node [left] {$P$} -- (0,0) -- node [above] {$P_s$} (4,0) -- (5,2) node [right] {$\frac{1}{4\beta}$};
\draw (-1,-2) node [left] {$\frac{1}{4\beta}$} -- (0,0);
\draw (4,0) -- (5,-2) node [right] {$\frac{1}{4\beta}$};
\node at (2, -4) {$(\beta)$};
\node at (-6, 0) {$\propto$};
\begin{scope}[shift = {(-18, 0)}]
\draw (0,0) node[left] {$\sqrt{2}P$} -- (3, 0) to [out = 60, in = 90] (7,0) node[right] {$\frac{P_s}{\sqrt{2}}$} to [out = -90, in = -60] (3,0);
\node at (3.5, -4) {$(\sqrt{2}\beta)$};
\end{scope}
 \end{tikzpicture}
\end{align}
So, if we prove crossing symmetry of Liouville theory with $c\leq 1$, it will follow that the theory is consistent not only on the sphere, but also on the torus and therefore on all oriented Riemann surfaces. 

\subsection{Why timelike Liouville theory does not exist}

Let us discuss the spectrum of Liouville theory with $c\leq 1$. 
In contrast to the case $c>1$, there is no continuum of unitary representations from which to build the spectrum. 
So we do not have any guidance from unitarity. 

However, we already know the three-point structure constant $C^{c\leq 1}$ and conformal blocks. 
In the $s$-channel decomposition of the four-point function,
\begin{align}
 \left< \prod_{i=1}^4 V_{P_i}(z_i) \right> 
= \int_{E} dP_s\ C^{c\leq 1}(P_1,P_2,P_s) C^{c\leq 1}(-P_s,P_3,P_4)\, 
\Big|\mathcal{F}^{(s)}_{P_s}(P_i|z_i)\Big|^2\ ,
\label{zfour}
\end{align}
the only unknown ingredient is the domain $E$ where the momentum $P_s$ should be integrated. 
Let us discuss what this domain can be, based on the properties of the integrand.
The integrand is a locally analytic function of $P_s$. 
It has poles from the conformal blocks, which are given by the general formula \eqref{polesblocks}.
For $c\leq 1$, that formula amounts to  
\begin{align}
 P_s \in \bigcup_\pm\, \pm \frac12\Big(\beta(1+\mathbb{N}) - \beta^{-1}(1+\mathbb{N})\Big)\ . 
\label{polesbr}
\end{align}
If $\beta^2\neq \mathbb{Q}$, then these poles are dense in the real line. 
There are also poles from the structure constants $C^{c\leq 1}$, which in contrast to the poles \eqref{polessc} from $C^\text{DOZZ}$ do not depend on the values of $P_1, \cdots P_4$. 
Instead, these poles are again on the real line,
\begin{align}
 P_s \in \bigcup_\pm \, \pm \frac12\Big(\beta\mathbb{N} + \beta^{-1}\mathbb{N}\Big) \, -\{0\}\ .
\label{polesscr}
\end{align}

So all the poles of the integrand are on the real line.
But the real line is where we should integrate $P_s$, if we assumed that the spectrum of Liouville theory was given for $c\leq 1$ by the same expression \eqref{alin} as for the rest of the values of $c$. 
It was proposed that for $c\leq 1$ there exists a timelike Liouville theory instead \cite{st03}, where the momentum $P_s$ would take imaginary values:
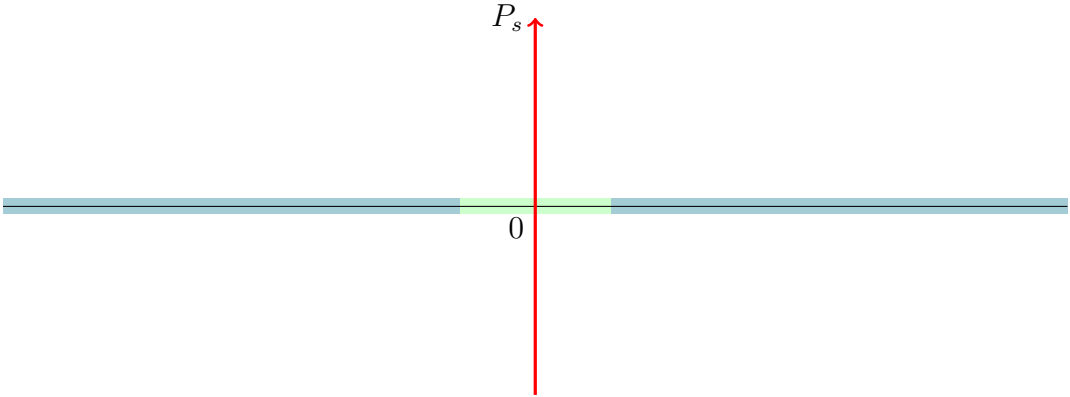
\begin{center}
 \begin{tikzpicture}
\fill [green, opacity = .2] (-7,-.1) rectangle (7,.1);
\fill [blue, opacity = .2] (-7,-.1) rectangle (-1, .1);
\fill [blue, opacity = .2] (1, -.1) rectangle (7, .1);
\draw[black] (-7, 0) -- (7,0);
\draw[red, very thick, ->] (0 ,-2.5) -- (0, 2.5) node[left, black]{$P_s$};
\node at (0,0) [below left] {$0$};
 \end{tikzpicture}
\end{center}
This proposal however has to be discarded, because the integral in the four-point function \eqref{zfour} diverges at $P_s = \pm i\infty$. 
Using the behaviour of the conformal blocks \eqref{dti}, and that of the three-point structure constants, which follows from eq. \eqref{lupi}, 
we indeed find
\begin{align}
 C^{c\leq 1}(P_1,P_2,P_s)C^{c\leq 1}(-P_s,P_3,P_4) \left|\mathcal{F}^{(s)}_{P_s}(P_i|z_i)\right|^2 \underset{P_s\to \pm i\infty}{\sim} |q|^{2P_s^2}\ ,
\end{align}
which is formally the same behaviour as for generic values of $c$ with $P_s\to \pm\infty$ \eqref{pinf}. 
So, in timelike Liouville theory, it is impossible to compute four-point functions. 

One might worry that our argument condemns all timelike theories. 
After all, the divergent factor $|q|^{2P^2_s}$ is universal, as it comes from the conformal blocks. 
However, the integral expression for the four-point function diverges only if the $s$-channel momentum $P_s$ is allowed to explore large imaginary values. 
So no divergences occur in the following models:
\begin{itemize}
\item \textbf{Timelike free bosonic theories:} Momentum conservation dictates the value of the $s$-channel momentum.
\item \textbf{$\widetilde{SL}_2(\mathbb{R})$ WZW model:} In this model (reviewed in \cite{rib14}), which describes strings in $AdS_3$, 
large negative conformal dimensions require large values of the spectral flow number, which is not quite conserved but can take at most three values in the $s$-channel.
\end{itemize}
So timelike theories can be fine, provided $s$-channel conformal dimensions are bounded from below in any given four-point function.
This can occur if interactions impose restrictions on $s$-channel quantum numbers. 
But there is no such restriction in timelike Liouville theory.

\subsection{Proposal for the spectrum and correlation functions}\label{sec:cscf}

We now explain our proposal for the spectrum of Liouville theory with $c\leq 1$.
In principle we keep the same real values for the momentum as for other values of $c$, but for computing four-point functions we  
move the $s$-channel momentum's integration line slightly away from the real line, in order to avoid all the poles. 
So the domain of integration in \eqref{zfour} is 
\begin{align}
 E= \mathbb{R} + i\epsilon\ ,
\end{align}
for some real number $\epsilon$:
\begin{center}
 \begin{tikzpicture}
\draw[red, very thick, ->] (-7 ,.6) -- (7,.6) node[above, black]{$P_s$};
\draw (0, -.1) -- (0, .1);
\fill [green, opacity = .2] (-7,-.1) rectangle (7,.1);
\fill [blue, opacity = .2] (-7,-.1) rectangle (-1, .1);
\fill [blue, opacity = .2] (1, -.1) rectangle (7, .1);
\draw[black] (-7, 0) -- (7,0);
\node at (0,0) [below] {$0$};
\draw[<->] (7.3, 0) -- (7.3, .6);
\node at (0,1.5){};
\node at (0,-.5){};
\node at (7.3,.3)[right] {$\epsilon$};
 \end{tikzpicture}
\end{center}
The four-point function does not depend on the value of $\epsilon$. 
On the one hand, the four-point function does not change when the integration line is deformed in the upper half-plane, so long the behaviour at infinity keeps $\int_E |q|^{2P_s^2}$ convergent. 
On the other hand, negative values of $\epsilon$ give the same result due to the invariance of the integrand under the reflection $P_s \to -P_s$. 

Then, instead of poles, the integrand of the four-point function has peaks:
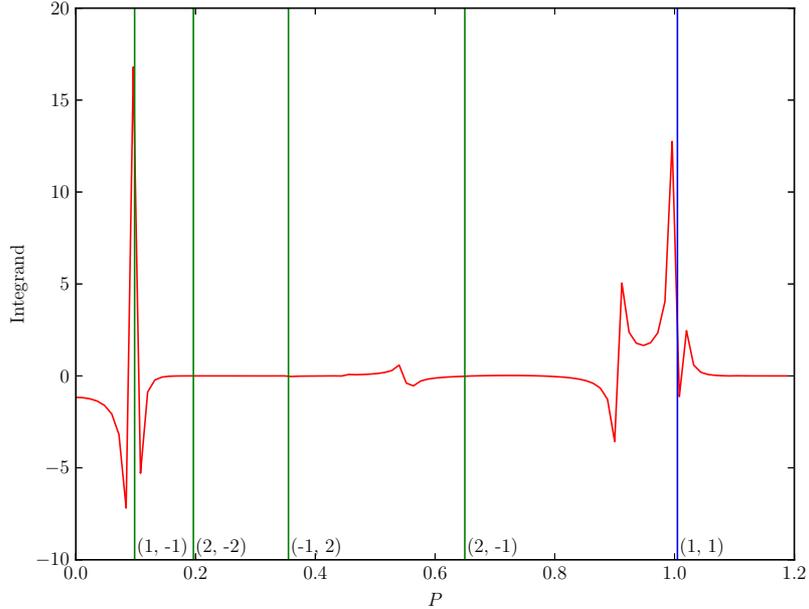
\begin{center}
\scalebox{.6}{\input{figure_integrand.pgf}}
\end{center}
Here we plotted the integrand as a function of $P=P_s-i\epsilon\in\mathbb{R}$. 
The parameters are $\beta = 1.103, (P_i) = (0.12, 0.7, 0.72, 0.95), (z_1,z_2, z_3, z_4) = (0.4, 0, \infty, 1)$, and $\epsilon = 3\times 10^{-3}$.
Some of the positions of the poles are indicated in terms of the integers $(m,n)$ such that $P = \frac12(m\beta + n\beta^{-1})$. 
Some poles with small residues do not produce noticeable peaks.

Crossing symmetry can now be checked numerically. 
For example, with $(P_i) = (0.32, 0.71, .45, .22)$ and $(z_i) = (0.27, 0, \infty, 1)$, we find
\begin{align}
 \begin{array}{|r|r|r|r|} \hline c & s\text{-channel} & t\text{-channel} & \text{precision} \\ \hline  \hline 0.873 & -3.68979 & -3.68975 & 1 \times 10^{-5} \\ \hline 0.564 & -3.17527 & -3.17526 & 3.3 \times 10^{-6} \\ \hline 0.241 & -2.74031 & -2.74032 & 2.4 \times 10^{-6} \\ \hline -1.237 & -1.49777 & -1.49778 & 1.1 \times 10^{-5} \\ \hline -3.751 & -0.50862 & -0.50861 & 2.6 \times 10^{-6} \\ \hline  \end{array} 
\end{align}
Again, the precision can be improved by increasing the numerical accuracy of the computations.
We can also study how the four-point function depends on the position $x$ with $(z_i) = (x, 0, \infty, 1)$. 
Taking $\beta = 1.103$ and $(P_i) = (0.18, -0.2, 0.71, -0.43)$, we find the following plot:
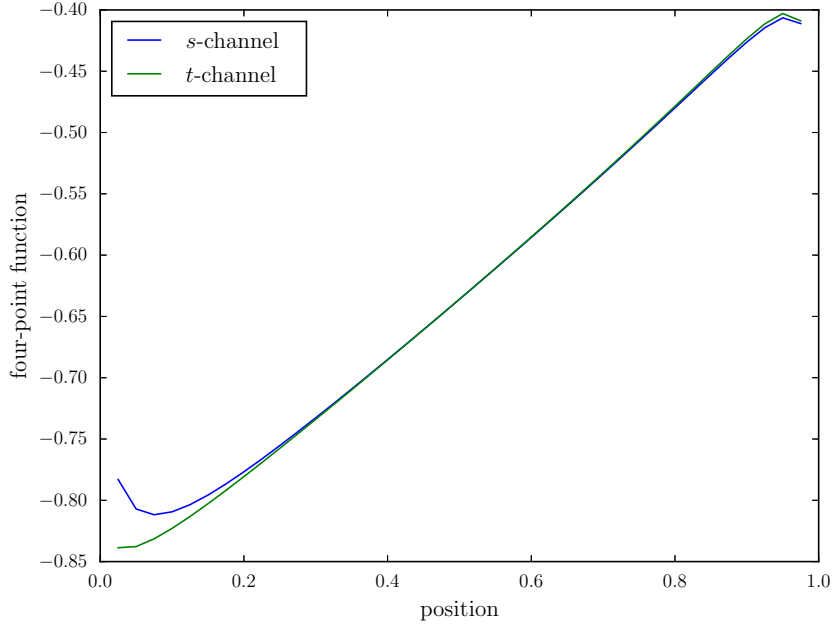
\begin{center}
 \scalebox{.6}{\input{figure_fourpoint.pgf}}
\end{center}
We did relatively low precision calculations, in order for some difference between the $s$- and $t$-channel results to be visible. 
The difference is visible near $x=0$ and $x=1$, where the series representations of the $t$- and $s$-channel conformal blocks reach their respective radiuses of convergence. 
With our specific choice of parameters, the difference happens to be more visible near $x=0$ than near $x=1$.

Let us discuss the continuation of this four-point function to other values of $c$ -- that is, complex values of $\beta$. 
As soon as  $\beta$ acquires a non-zero imaginary part, the poles of the integrand leave the real line, and infinitely many poles cross our integration line:
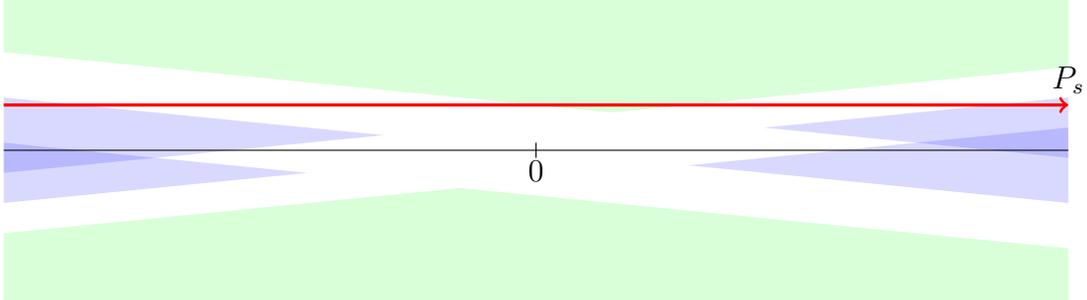
\begin{center}
 \begin{tikzpicture}
\fill[opacity = .15, blue] (3,.3) -- (7,.7) -- (7, -.1) -- cycle;
\fill[opacity = .15, blue] (2, -.2) -- (7, .3) -- (7, -.7) -- cycle;
\fill[opacity = .15, blue] (-3,-.3) -- (-7,-.7) -- (-7, .1) -- cycle;
\fill[opacity = .15, blue] (-2, .2) -- (-7, -.3) -- (-7, .7) -- cycle;
\fill[opacity = .15, green] (1,.5) -- (7, 1.1) -- (7, 2) -- (-7, 2) -- (-7, 1.3) -- cycle;
\fill[opacity = .15, green] (-1,-.5) -- (-7, -1.1) -- (-7, -2) -- (7, -2) -- (7, -1.3) -- cycle;
\draw[red, very thick, ->] (-7 ,.6) -- (7,.6) node[above, black]{$P_s$};
\draw (0, -.1) -- (0, .1);
\draw[black] (-7, 0) -- (7,0);
\node at (0,0) [below] {$0$};
 \end{tikzpicture}
\end{center}
So although its integrand is analytic in $\beta$, the four-point function \eqref{zfour} does not have an analytic continuation to complex values of $\beta$.
Actually, the existence of such an analytic continuation would have been an embarrassment of riches, as it would have provided a second definition of Liouville theory for $c\in \mathbb{C}$. 
On the other hand, the correlation functions are analytic in the momentums $P_i$, and can be computed for $P_i\in \mathbb{C}$ and not just $P_i \in \mathbb{R}$. 

Finally, let us study the behaviour of the four-point function near $z_1=z_2$.
The universal behaviour of four-point conformal blocks is
\begin{align}
 \mathcal{F}^{(s)}_{P_s}(P_i|z_i) \underset{z_1\to z_2}{\sim} |z_{12}|^{\delta + 2P_s^2} \quad \text{where} \quad \delta = -2P_1^2-2P_2^2-\frac{Q^2}{2}\ .
\end{align}
On the other hand, 
$C^{\text{DOZZ}}(P_1,P_2,P_3)$ has a simple zero at $P_1=0$ while $C^{c\leq 1}(P_1,P_2,P_3)$ has a finite limit.
We must distinguish two cases:
\begin{alignat}{3}
 \left< \prod_{i=1}^4 V_{P_i}(z_i) \right> 
& \underset{z_1\to z_2}{\sim}  \int_\mathbb{R} dP_s\ |z_{12}|^{\delta + 2P_s^2} 
&& \sim |z_{12}|^\delta |\log|z_{12}||^{-\frac12}\ ,
&& \qquad (c\leq 1)
\\
& \underset{z_1\to z_2}{\sim} \int_\mathbb{R} dP_s\ |z_{12}|^{\delta + 2P_s^2} P_s^2 
&& \sim |z_{12}|^\delta |\log|z_{12}||^{-\frac32}\ . 
&& \qquad (\text{otherwise})
\end{alignat}
Therefore, the leading behaviour of the four-point function near $z_1=z_2$ depends on whether $c\leq 1$ or not.
In both cases there are corrections in powers of $z_{12}$ and $\frac{1}{\log|z_{12}|}$.
The logarithms which appear here are due to the continuous spectrum, and have nothing to do with the logarithms of logarithmic conformal theory or of Coulomb gas correlation functions \cite{savi13}.

\section{Non-analytic theories at rational values of \texorpdfstring{$c$}{c}}

In our discussion of Liouville theory, we assumed that correlation functions are analytic functions of the central charge $c$ and of the momentums of the fields. 
Under this assumption, and some other assumptions, there exists a unique non-rational conformal field theory at any complex value of $c$, which we proposed to call Liouville theory. 

We will now see that relaxing the analyticity assumption allows other theories to exist, when $c$ takes the discrete values of eq. \eqref{cpq} -- which we will call rational values. 
We give these theories the generic name of non-analytic Liouville theory, as their correlation functions are analytic neither in $c$ nor in the momentums.
The prototype of these theories is Runkel--Watts theory at $c=1$ \cite{rw01}. 
The generalization to other discrete values of $c$ was proposed by McElgin \cite{mce07}. 

We will check that McElgin's proposal obeys crossing symmetry.
We will also rule out other plausible proposals, including a puzzling family of three-point functions which obey crossing symmetry for some but not all values of the momentums.

\subsection{Non-analytic Liouville theory}

As we discussed in Section \ref{sec:scse}, the degenerate crossing symmetry equations have a unique solution provided $b$ and $b^{-1}$ are aligned and incommensurable complex numbers. 
If however $\beta^2 = \frac{q}{p}\in\mathbb{Q}$, then there exist smooth functions which are periodic with both periods $b$ and $b^{-1}$.
Using such functions, it is possible to build infinitely many three-point structure constants which solve the degenerate crossing symmetry equations. 
Of course, we do not expect all these solutions to solve the full crossing symmetry equations. 
This nevertheless suggests that for $\beta^2=\frac{q}{p}$ there may exist alternative solutions.

An even more suggestive observation is that the DOZZ three-point structure constant, while it does not have a limit for generic values of $c\leq 1$, does have a limit for rational values of $c$. 
This was first observed in the case $c=1$ \cite{sch03}, where the limit of the DOZZ structure constant was found to agree with the already known structure constant of Runkel--Watts theory. 
This was then generalized to the other rational central charges \cite{mce07}.  
The resulting three-point structure constant is 
\begin{align}
 C^\text{non-analytic}(P_1,P_2,P_3) = \lim_{\beta^2 \to \frac{q}{p}} C^{\text{DOZZ}}(P_1, P_2, P_3) =  C^{c\leq 1}(P_1,P_2,P_3) \sigma(P_1, P_2, P_3)\ , 
\end{align}
where we define, assuming that $p$ and $q$ are coprime integers,
\begin{align}
\sigma(P_1, P_2, P_3) 
&= \left\{\begin{array}{l} 
	  1 \quad \text{if} \quad \prod_{\pm, \pm} \sin\pi\Big(\frac12(p-q)+\sqrt{pq}(P_1\pm P_2\pm P_3)\Big) < 0 \ ,
          \\
	  0 \quad \text{else}\ .
         \end{array} \right.
\label{sigma}
\end{align}
This step function is the limit of $\frac{C^{\text{DOZZ}}}{C^{c\leq 1}}$, which is defined for $\beta^2\notin\mathbb{R}$ and invariant under shifts of momentums by $b$ and $b^{-1}$. 
We will now plot $C^{\text{non-analytic}}(P_1, P_2, P_3)$ and $C^{c\leq 1}(P_1, P_2, P_3)$ as functions of $P_1$, and  draw green vertical lines at $P_1\in \frac{1}{2\sqrt{pq}}\mathbb{Z}$, where $\sigma(P_1, P_2, P_3)$ always vanishes. 
We take $(p, q) = (7, 5)$ and $(P_2, P_3) = (0.2, 0.45)$, and smooth the poles of $C^{c\leq 1}$ into peaks by a small shift $P_1\to P_1+i\epsilon$: 
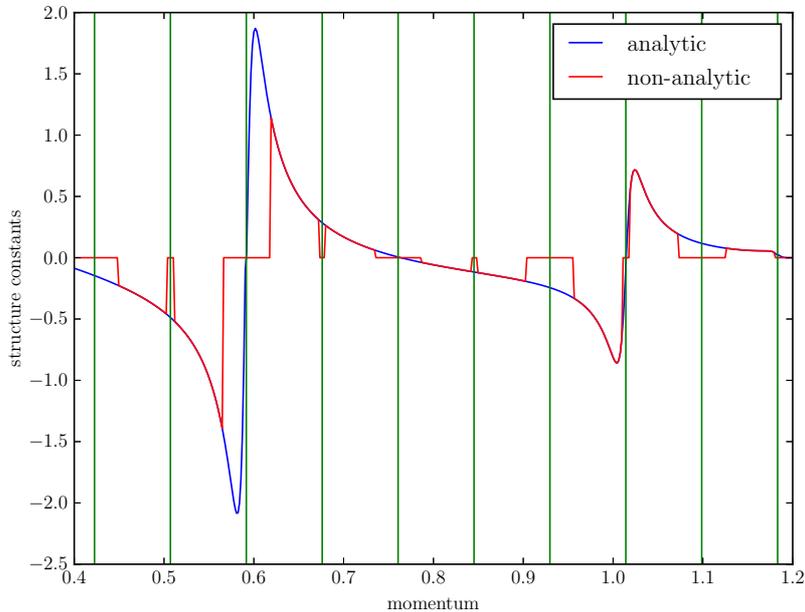
\begin{center}
\scalebox{.6}{\input{figure_non_analytic.pgf}}
\end{center}
We now consider the $s$-channel four-point function which corresponds to $C^\text{non-analytic}$, 
\begin{align}
 \left< \prod_{i=1}^4 V_{P_i}(z_i) \right> 
= \int_{\substack{\sigma(P_1, P_2, P_s)=1 \\ \sigma(-P_s, P_3, P_4) = 1}} dP_s\ C^{c\leq 1}(P_1,P_2,P_s) C^{c\leq 1}(-P_s,P_3,P_4)\, 
\Big|\mathcal{F}^{(s)}_{P_s}(P_i|z_i)\Big|^2\ .
\end{align}
We do not need to shift the integration contour away from the real line, as we did in Section \ref{sec:cscf}, because the poles \eqref{polesbr} and \eqref{polesscr} of the integrand all belong to $\frac12(\beta \mathbb{Z} + \beta^{-1}\mathbb{Z}) = \frac{1}{2\sqrt{pq}}\mathbb{Z}$, and are therefore outside the integration domain. 
This implies that our four-point function is a limit of the Liouville four-point function at complex central charges, as taking this limit does not involve poles crossing the integration domain. 
This in turn implies that non-analytic Liouville theory is crossing-symmetric.

Near rational values of the central charge, Liouville correlation functions vary violently as functions of momentums. 
This makes it difficult to numerically check that their limits are indeed the correlation functions of non-analytic Liouville theory.
So we focused on directly checking that non-analytic Liouville theory is crossing-symmetric.
For example, with $(z_i) = (0.23, 0, \infty, 1)$ and $(P_i) = (0.22, 0.37, 0.28, 0.12)$, 
we find 
\begin{align}
\renewcommand*{\arraystretch}{1.1}
\begin{array}{|r|r|r|r|} \hline c & s\text{-channel} & t\text{-channel} & \text{precision} \\ \hline  \hline - \frac{25}{2} & 0.0245541 & 0.0245559 & 7.7 \times 10^{-5} \\ \hline - \frac{3}{5} & 0.0547238 & 0.0547235 & 5.5 \times 10^{-6} \\ \hline 0 & 0.0361717 & 0.0361713 & 1.1 \times 10^{-5} \\ \hline \frac{1}{2} & 0.0909525 & 0.0909523 & 6.7 \times 10^{-6} \\ \hline 1 & 0.2937675 & 0.2937677 & 6.4 \times 10^{-7} \\ \hline  \end{array} 
\end{align}

\subsection{Other plausible theories}

Let us relax the assumption that $p$ and $q$ are coprime in the definition of the non-analytic factor $\sigma$ \eqref{sigma}. 
For each rational value of the central charge, we then obtain infinitely many conjectural three-point structure constants, parametrized by pairs of integers $(p,q)$ with a fixed ratio. 
Crossing symmetry is no longer guaranteed, because the corresponding correlation functions are no longer limits of Liouville correlation functions at complex central charges. 
And we numerically find that crossing symmetry is not obeyed. 

Another idea is to replace the non-analytic factor $\sigma$ with any function of $\sigma$, as was suggested in \cite{er13} in the case $c=1$. Since $\sigma$ takes only two values, this amounts to considering three-point functions of the type
\begin{align}
 C_\nu (P_1,P_2,P_3) &= C^{c\leq 1}(P_1,P_2,P_3) \Big((1-\nu) +\nu \sigma(P_1, P_2, P_3) \Big)\ , 
\end{align}
for some parameter $\nu$, with $C_0 = C^{c\leq 1}$ and $C_1 = C^\text{non-analytic}$.
The $s$-channel decomposition of the four-point function reads 
\begin{multline}
 \left< \prod_{i=1}^4 V_{P_i}(z_i) \right> 
= \left( (1-\nu)^2 \int_{\mathbb{R} + i\epsilon } + \nu(1-\nu) \int_{E\sqcup E'} + \nu(2-\nu) \int_{E\cap E'} \right) 
dP_s
\\ 
C^{c\leq 1}(P_1,P_2,P_s) C^{c\leq 1}(-P_s,P_3,P_4)\, 
\Big|\mathcal{F}^{(s)}_{P_s}(P_i|z_i)\Big|^2\ ,
\label{tmn}
\end{multline}
which involves the disjoint union and intersection of the sets $E$ and $E'$ defined by
\begin{align}
 \mathbf{1}_E(P_s) & = \sigma(P_1, P_2, P_s)\ ,
\\
\mathbf{1}_{E'}(P_s) &= \sigma(-P_s, P_3, P_4)\ .
\end{align}
(We used the identities $\mathbf{1}_E\mathbf{1}_{E'}=\mathbf{1}_{E\cap E'}$ and $\mathbf{1}_E+\mathbf{1}_{E'}=2\cdot\mathbf{1}_{E\cap E'} + \mathbf{1}_{E\sqcup E'}$.)
As functions of $\nu$, the three terms of the four-point function are independent polynomials, and we can investigate crossing symmetry term by term. 
The first and third term correspond to Liouville theory and non-analytic Liouville theory, and are therefore crossing-symmetric. 
It remains to investigate crossing symmetry of the second term -- the integral over the disjoint union $E\sqcup E'$.
Curiously, we numerically find that the second term can be crossing symmetric or not, depending on the values of $P_1, \cdots P_4$. 
For example, taking $(P_1, P_2, P_3) = (0.22, 0.47, 0.31)$ with $c=1$ and $(z_i) = (0.23, 0, \infty, 1)$, we find the following behaviour of the $s$- and $t$-channel integrals as functions of $P_4$:
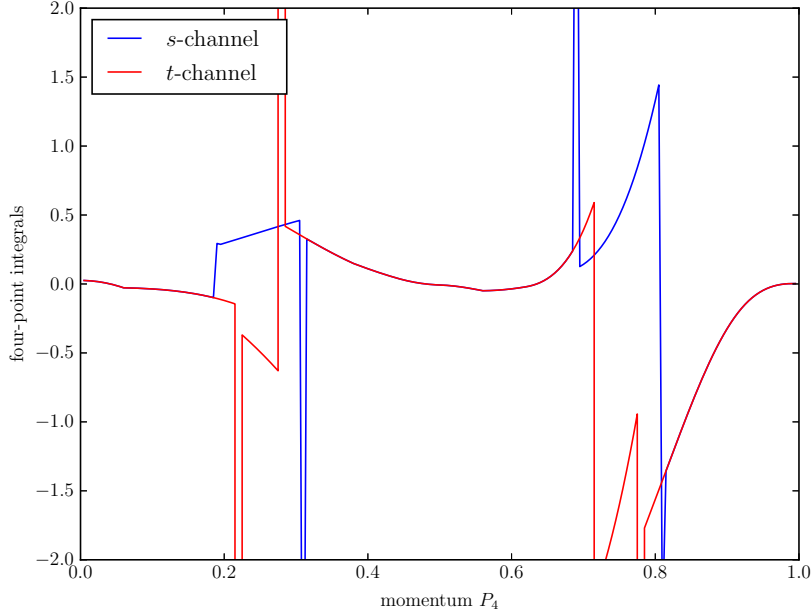
\begin{center}
\scalebox{.6}{\input{figure_disjoint.pgf}}
\end{center}
Based on this and other examples, we conjecture that crossing symmetry is obeyed if
\begin{align}
 \Big((p_1, p_3)\cup (p_3, p_1)\Big) \cap \Big((p_2, p_4)\cup (p_4, p_2)\Big) = \emptyset \ , 
\ \text{with}\  
p_i = \pm \sqrt{pq} P_i\ \operatorname{mod}\ \tfrac{1}{2} \in \left(0, \tfrac{1}{4}\right)\ .
\label{pi}
\end{align}
This condition implies that the $s$-channel integration domain $E\sqcup E'$ is made of intervals of the same lengths as the $t$-channel integration domain. 

In principle, crossing symmetry means the agreement of three decompositions, corresponding to the $s$, $t$ and $u$ channels. 
So far we have only considered the $s$ and $t$ channels, whose agreement for all values of the momentums implies that they also agree with the $u$ channel. 
Now, in the case of $C_\nu$, we see that for any choice of the four momentums, two of the three channels agree. 
The third channel corresponds to using the operator product expansion of the two fields with the lowest values of $p_i$. 
This observation would deserve to be further checked, understood, and generalized to $n$-point functions.

\section{Conclusion}

Our results complete the definition of Liouville theory for any complex value of the central charge $c$. 
The spectrum is always given by the formula \eqref{alin}.
On the other hand, the formula for the three-point structure constant depends on whether $c\leq 1$ \eqref{cleq1} or not \eqref{cdozz}.

Liouville theory can be characterized as the unique two-dimensional conformal field theory whose spectrum is continuous with multiplicities one, and whose correlation functions are smooth as functions of the central charge and momentums.
Relaxing some of these conditions, we obtain other theories. 
In particular, we called non-analytic Liouville theory the theory which is obtained for rational values of $c\leq 1$ \eqref{cpq} as the limit of Liouville theory with complex $c$. 

We have also performed numerical checks of crossing symmetry in the generalized minimal models, which exist at all values of $c$ with the discrete spectrum $\mathcal{S} = \bigoplus_{r,s=1}^\infty \mathcal{R}_{(r,s)}\otimes \bar{\mathcal{R}}_{(r,s)}$.
In contrast to Liouville theory, generalized minimal models exist only on the sphere and not on higher genus Riemann surfaces, according to the following heuristic argument. 
The torus partition functions of generalized minimal models and Liouville theory are both infinite. 
In Liouville theory, torus correlation functions become finite as soon as non-degenerate primary fields are present. In generalized minimal models, all fields are degenerate and their torus correlation functions remain infinite -- except when they vanish due to the fusion rules.

Now that we have a consistent non-rational conformal field theory for any value of $c\leq 1$, we can come back to our original motivation of accounting for statistical physics observables. 
Our results provide well-defined predictions for correlation functions, which can be 
numerically computed to high accuracy for any values of the conformal dimensions and of the central charge. 
The challenge is to find matching observables in statistical systems. 
For example, it would be interesting to compare our results with four-spins correlation functions in the $Q$-states Potts model for general values of $Q$.
Our results may also have applications in four-dimensional gauge theories, in the context of their relation with Liouville theory with a central charge less than one \cite{shsu12,bbrt14}.

\appendix

\section{Numerical calculations}\label{sec:nccs}

In this Appendix we discuss which expressions for conformal blocks and correlation functions are suitable for numerical calculations, and which cutoffs and approximations are used. 
We do not specifically describe our program for doing the calculations, but focus on the underlying principles and formulas.

\subsection{Four-point functions}

An $s$-channel four-point function is an integral over an $s$-channel momentum $P_s$, see for example eq. \eqref{zfour} in the case of Liouville theory with $c\leq 1$.
In Liouville theory, the integration domain is $P_s\in\mathbb{R}+i\epsilon$ for $c\leq 1$, and $P_s\in \mathbb{R}$ otherwise.
In the latter case we can use the reflection symmetry $P_s\to -P_s$ of the integrand, and halve the integration domain to 
$P_s \in (0,\infty)$.

We introduce a cutoff $P_\text{max}$ for the integral on $P_s$. 
Since the integral is exponentially convergent near $P_s=\infty$, it is not necessary to take large values of $P_\text{max}$. 
In the case $c\leq 1$ we have the additional parameter $\epsilon$, and the integral is over $(-P_\text{max} + i\epsilon, P_\text{max} + i\epsilon)$. 
Four-point functions should not depend on $\epsilon$ at all. 
In practice $\epsilon$ should be large enough for the integration line to be at a safe distance from the poles, and small enough for the integrals over the segments $(P_\text{max}, P_\text{max} + i\epsilon)$ and $(-P_\text{max}, -P_\text{max} + i\epsilon)$ to be negligible. 
We find that the four-point function is indeed constant over a wide range of values of $\epsilon$, typically $O(10^{-2}) - O(1)$.

Our goal of computing correlation functions influences the way we compute structure constants and conformal blocks:
\begin{itemize}
 \item Conformal blocks are rational functions of $P_s$, 
with poles at the degenerate values $P_s = P_{(r,s)}$ for $r,s\in \mathbb{N}^*$. 
We should compute the residues at these poles once and for all, in order to avoid recomputing them for each value of $P_s$ when integrating over $P_s$. 
\item Structure constants depend neither on $z_i$, nor the conformal blocks' cutoffs (such as the cutoff on $(r,s)$ when summing over poles). 
We should avoid recomputing the structure constants when changing the values of $z_i$ or these cutoffs.
We however use built-in methods for integrating over $P_s$, and we do not control which values of $P_s$ are used. 
This problem is solved by using spline interpolations of the structure constants as functions of $P_s$. 
\end{itemize}
 
\subsection{Conformal blocks}\label{sec:ccb}

Global conformal symmetry determines the dependences of four-point blocks on three of the four positions $z_i$, and we assume $(z_i) = (x, 0, \infty, 1)$. 
Let us write the conformal blocks in a way which is convenient for our numerical calculations,
\begin{multline}
 \mathcal{F}^{(s)}_{P_s}(P_i|x,0,\infty, 1) 
=   x^{-\frac{Q^2}{4}-P_1^2-P^2_2} (1-x)^{-\frac{Q^2}{4}-P^2_1-P^2_4} 
\\
\times
 (16q)^{P_s^2}\theta_3(q)^{-Q^2-4(P_1^2+P_2^2+P_3^2+P_4^2)} \left(1 + \sum_{N=1}^{\infty} \sum_{rs\leq N} C^N_{(r,s)} \frac{(16q)^N}{P_s^2 - P_{(r,s)}^2}\right)\ ,
\label{block}
\end{multline}
where we introduced the elliptic variable $q$ and function $\theta_3(q)$,
\begin{align}
 q = \exp -\pi \frac{F(\frac12,\frac12,1,1-x)}{F(\frac12,\frac12,1,x)}  \quad , \quad \theta_3(q) = \sum_{n\in{\mathbb{Z}}} q^{n^2}\ .
\end{align}
The coefficients $C^N_{(r,s)}$ are given by the recursion relation 
\begin{align}
 C^N_{(r,s)} = R_{(r,s)} \left(\delta_{N-rs,0} + \sum_{r's'\leq N- rs} \frac{C^{N-rs}_{(r',s')}}{P^2_{(r,-s)} - P^2_{(r',s')}} \right)\ ,
\label{cb: CN}
\end{align}
where the prefactor is 
\begin{align}
 R_{(r,s)} = \frac{-2P_{( 0,0)} P_{( r,s)}}{\prod_{r'=1-r}^r \prod_{s'=1-s}^s 2P_{( r',s')}}
\prod_{r'\overset{2}{=}1-r}^{r-1} \prod_{s'\overset{2}{=}1-s}^{s-1} \prod_\pm (P_2\pm P_1 + P_{( r',s')}) (P_3\pm P_4 +P_{( r',s')})\ .
\label{cb: Rnm}
\end{align}
In numerical calculations, we introduce a cutoff $N_\text{max}$ on the summation index $N$ in eq. \eqref{block}. 
We find that increasing $N_\text{max}$ by one often does not improve accuracy if $N_\text{max}$ is even, so we take only even values of $N_\text{max}$. 
In this approximation,
the non-trivial factor of the conformal block is a polynomial function of $q$ of degree $N_\text{max}$, and a rational function of $P_s^2$ with the poles $\{P_{(r,s)}^2\}_{rs\leq N_\text{max}}$. 

Our formulas for conformal blocks are a reformulation of Al. Zamolodchikov's recursion representation. (See \cite{rib14} for a review.) Recursive formulas are particularly efficient for numerical computations, as compared to combinatorial formulas. 
The recursion representation which we used expresses a conformal block as series in $q$, whose radius of convergence is one. 
With the elliptic variable $q$, the duality which relates $s$-channel blocks with a position $x$ and $t$-channel blocks with the position $1-x$ has the self-dual value $q_\text{self-dual}=e^{-\pi} \simeq 0.0432$.
Since this value is small, there exists a regime where both $s$- and $t$-channel decompositions of the same four-point function involve conformal blocks with quickly convergent series expansions.

\subsection{Structure constants}\label{sec:csc}

The Liouville structure constant has different expressions $C^{c\leq 1}$ \eqref{cleq1} and $C^\text{DOZZ}$ \eqref{cdozz}, depending on whether $c\leq 1$. 
Both expressions are products of $\Upsilon_b$ functions, but the arguments of these functions behave very differently.
In the case of $C^{c\leq 1}$, the arguments can have arbitrarily large real parts. 
So we can use the integral formula \eqref{lup} for the $\Upsilon_b$ functions only after repeated use of the shift equations \eqref{upup}, and actually it is simpler to use the product formula \eqref{upp} instead. 
In the case of $C^\text{DOZZ}$, the arguments have real parts $\Re Q $ or $\frac12 \Re{Q}$.
So we can more easily use the integral formula, which converges faster than the product formula.

Let us start with the case of $C^\text{DOZZ}$. 
In the integrand of the four-point function, we have a factor $\prod_\pm \Upsilon_b(Q\pm 2iP_s)$, where $Q\pm 2iP_s$ are at the edge of the domain where the integral formula for $\Upsilon_b$ converges.
Using the shift equations, we find
\begin{align}
 \prod_\pm \Upsilon_b (Q\pm 2iP_s) = 4P_s^2 \prod_\pm \Upsilon_b(b\pm 2iP_s)\ .
\end{align}
Using the integral formula, we now find 
\begin{align}
 \prod_\pm \frac{\Upsilon_b (Q\pm 2iP_s)}{\Upsilon_b (\frac{Q}{2}\pm 2iP_s)} = 4\Upsilon_b(b)^2 P_s^2  \exp \int_0^\infty \frac{dt}{t}\frac{4\sin^22P_st\sinh^2\frac12 (b-\frac{1}{b})t}{\sinh bt \sinh\frac{t}{b}}\ .
\end{align}
We will also use the identity
\begin{multline}
 \log \frac{\prod_{i=1}^3 \Upsilon_b(\frac{Q}{2}+2iP_i)}{\prod_{\pm,\pm}\Upsilon_b(\frac{Q}{2}+iP_1\pm iP_2 \pm iP_3)} 
\\
= -4\int_0^\infty\frac{dt}{t} \frac{\sum_{i=1}^3\sin^4 P_it -2\sum_{i<j}\sin^2P_it\sin^2P_jt + 4\prod_{i=1}^3\sin^2P_it}{\sinh bt\sinh\frac{t}{b}}\ ,
\end{multline}
which follows from trigonometric manipulations, using in particular 
\begin{align}
 \sum_{\pm,\cdots \pm}\sin^2(a_1\pm a_2 \cdots \pm a_n) = 2^{n-2}\left(1-\prod_{i=1}^n\cos 2a_i\right)\ .
\end{align}
The $P_s$-dependence of the product of DOZZ structure constants which appears in a four-point function can therefore be written in terms of a single convergent integral, and we have
\begin{multline}
 C^\text{DOZZ}(P_1,P_2,P_s) C^\text{DOZZ}(-P_s, P_3, P_4)
 = 4\Upsilon_b(b)^2 \prod_{i=1}^4\frac{\Upsilon_b(Q+2iP_i)}{\Upsilon_b(\frac{Q}{2}+2iP_i)}
\\
\times P_s^2 \exp \int_0^\infty \frac{dt}{t} \frac{4\varphi(t)}{\sinh bt\sinh\frac{t}{b}}\ , 
\end{multline}
where we introduce the function
\begin{multline}
 \varphi(t) = \sin^2 2P_st\sinh^2 \tfrac12(b-\tfrac{1}{b})t-2\sin^4 P_st-(\sin^2P_1t-\sin^2P_2t)^2-(\sin^2P_3t-\sin^2P_4t)^2
\\
+\sin^2 P_st\left(2\sum_{i=1}^4\sin^2P_it-4\sin^2P_1t\sin^2P_2t-4\sin^2P_3t\sin^2P_4t\right)\ .
\end{multline}
If we instead used the integral formula for each $\Upsilon_b$ factor with a $P_s$-dependent argument, we would have to perform $10$ integrals, which converge at $t=0$ only thanks to regularizing terms in the integrands. 

In the case of $C^{c\leq 1}$, the problem is the slow convergence of the infinite product \eqref{upp}, which is not very well approximated by finite products with moderate numbers of factors. 
This can however be improved  by approximating the infinitely many neglected factors in terms of an integral. 
To do this, we rewrite our infinite product as 
\begin{align}
 \Upsilon_b(\tfrac{Q}{2}+iP) = \lambda_b^{-P^2} \exp\sum_{m,n=0}^\infty \psi''\left(\frac{y_{m,n}}{P}\right)\ ,
\end{align}
where we introduce 
\begin{align}
 y_{m,n} = \frac{Q}{2} + mb+nb^{-1}\ ,
\end{align}
and the function 
\begin{align}
 \psi''(y) = \log \left(1+\frac{1}{y^2}\right) -\frac{1}{y^2}\ .
\end{align}
The idea is now to use the approximation
\begin{align}
 \sum_{m,n=0}^\infty \sim \sum_{m=0}^M \sum_{n=0}^N + \sum_{m=0}^M \int_{N+\frac12}^\infty dn +\sum_{n=0}^N \int_{M+\frac12}^\infty dm + \int_{M+\frac12}^\infty\int_{N+\frac12}^\infty dmdn\ .
\end{align}
This can be made explicit using the following primitives of $\psi''$, 
\begin{align}
\psi'(y) &= y\log\left(1+\frac{1}{y^2}\right) +\frac{1}{y} + i\log\frac{y+i}{y-i}\ ,
\\
 \psi(y) &= \frac32 + \frac12 (y^2-1)\log\left(1+\frac{1}{y^2}\right) + iy\log\frac{y+i}{y-i}\ ,
\end{align}
which are such that $\psi(y) \underset{y\to\infty}{=} O(\frac{1}{y^2})$.
We obtain the approximation
\begin{multline}
 \Upsilon_b(\tfrac{Q}{2}+iP) \sim \lambda_b^{-P^2} \prod_{m=0}^M \prod_{n=0}^N \left(1+\frac{P^2}{y_{m,n}^2}\right) 
\\
\times e^{ - \sum_{m=0}^M\sum_{n=0}^N \frac{P^2}{y_{m,n}^2} - Pb\sum_{m=0}^M \psi'\left(\frac{y_{m,N+\frac12}}{P}\right) - Pb^{-1}\sum_{n=0}^N\psi'\left(\frac{y_{M+\frac12,n}}{P}\right) + P^2 \psi\left(\frac{y_{M+\frac12,N+\frac12}}{P}\right) }\ .
\end{multline}
This converges acceptably fast, and allows us to accurately compute the values of $\Upsilon_b$ using relatively small values of the cutoffs $M$ and $N$.

\acknowledgments{
We wish to thank Olivier Babelon, Marco Picco, Volker Schomerus, and particularly Vladimir Belavin, for helpful comments on the draft of this article. We are grateful to Teresa Bautista Solansa, 
Atish Dabholkar, and Harold Erbin for intense discussions. R. S. is grateful to Jacopo Viti for interesting discussions, and to Vincent Degat for help with a computer. We thank Pasha Gavrylenko for reporting a mistake in eq. \eqref{tmn}.
}


\end{document}